# Unsaturated Dinitrogen Difluoride under Pressure: toward high-Energy Density Polymerized NF Chains

Guo Chen[1,2], Ling Lin[3], Chengfeng Zhang[1,2], Jie Zhang[1,*], Gilles Frapper[4] and Xianlong Wang[1,2*]

[1]*Key Laboratory of Materials Physics, Institute of Solid State Physics, HFIPS, Chinese Academy of Sciences, Hefei 230031, China*

[2]*University of Science and Technology of China, Hefei 230026, China*

[3]*State Key Laboratory of Fluorine and Nitrogen Chemistry and Advanced Materials，Shanghai Institute of Organic Chemistry, Chinese Academy of Sciences,Shanghai 200032, China*

[4]*Applied Quantum Chemistry group, E4, IC2MP, UMR 7285 Poitiers University, CNRS, 4 rue Michel Brunet TSA 51106, 86073 Poitiers Cedex 9, France.*

─────

[*]Author to whom all correspondence should be addressed: xlwang@theory.issp.ac.cn,

jzhang@theory.issp.ac.cn

**ABSTRACT**

Based on first-principles calculations and ab initio molecular dynamics simulations, the polymerisation of the unsaturated *cis* dinitrogen-difluoride (*cis*-$N_2F_2$) molecular compound is investigated. The thermodynamic, dynamical and thermal stabilities of the nitrogen fluorine NF system are investigated at conditions of 0-3000 K and 0-200 GPa. The *cis*-$N_2F_2$ molecule is a suitable precursor to obtain one-dimensional polymerized nitrogen-fluorine (poly-NF) chains at a pressure above 90 GPa and at a temperature around 1900 K. Importantly, these poly-NF chains can be quenched to room conditions, and potentially serve as a High-energy-density materials (HEDM). It has been established that when Al is utilised as a reducing agent, poly-NF chains exhibit a gravimetric energy density of 13.55 kJ/g, which exceeds that of cubic gauche nitrogen (cg-N, 9.70 kJ/g). This is attributable to the presence of both polymerised nitrogen and strong oxidising F atoms.

# INTRODUCTION

Nitrogen gas makes up about 78% of Earth's atmosphere, and one key area of high-pressure physics involves converting nitrogen gas into polymeric nitrogen, a material with high energy density (HEDM) [1]. This conversion relies on breaking the strong triple bond (N≡N, bonding energy of 954 kJ/mol) in nitrogen to form polymeric structures with single bond (N–N, bonding energy of 160 kJ/mol) and double bond (N=N, bonding energy of 418 kJ/mol) [2]. These bonds have significant energy differences, with the N≡N bond releasing a large amount of energy when dissociated. Polymeric nitrogen shows potential for use in energy storage, propellants, and explosives due to its high energy density.

Creating polymeric nitrogen is challenging because the N≡N bond is chemically inert under ambient pressure. High-pressure techniques can synthesize different polymeric nitrogen phases [e.g., cubic gauche nitrogen (cg-N), layered nitrogen, etc.] [3–7]. but these materials cannot be stabilized at ambient pressure. To overcome this, researchers are working to reduce the synthesis pressure and improve stability, employing strategies like surface modification [8], nano-confinement [9], and doping techniques [10]. For examples, the $N_8$ and cg-N have been observed to remain stable within carbon nanotubes under room conditions [11,12], and doping with elements like phosphorus has shown potential for stabilizing black phosphorus nitrogen (BP-N) [13].

Most studies have shown that the electronegativity of doping elements is generally lower than that of nitrogen [14]. In these studies, polymeric nitrogen carries a negative charge and forms covalent or metallic bonds with the introduced elements. However, if the doping element has a higher electronegativity than N element, the charge transfer from N to doping element can also lead to new polymer nitrogen patterns. For instance, using fluorine, the most electronegative element, greatly influences the compounds it forms [15,16]. However, although theoretical studies predicted the existence of certain nitrogen-fluorine (NF) compounds under high pressure, these compounds have not been successfully synthesized yet.

Recent research predicts that polymerized nitrogen-fluorine (poly-NF) chains in the *Cmca* phase can be formed under pressures of 90-200 GPa and could potentially release large amounts of energy when reacted with aluminum powder. The study uses first-principles calculations and ab initio Molecular Dynamics (AIMD) methods to explore the behavior of *cis* dinitrogen-difluoride (*cis*-$N_2F_2$) molecules at varying pressures and temperatures. These simulations suggest that poly-NF chains can be formed at high pressures and can be quenched to ambient conditions, offering a promising path for future applications.

**METHODS**

Structure relaxations (shape, volume, atomic positions) are performed according to the density functional theory within the framework of the all-electron projector augmented wave (PAW) method, as implemented in the VASP code (version 5.4.4) [17,18]. We adopt the Perdew−Burke−Ernzerhof (PBE) functional [19] at the generalized gradient approximation (GGA) [20,21] level of theory. Valence states of $2s^2p^3$ for N, $2s^22p^7$ for F, and $3s^2p^1$ for Al atoms were utilized. A plane-wave kinetic energy cutoff of 520 eV is used within a uniform.

All structures are optimized at pressures from 0 to 200 GPa until the net forces on atoms are below 1 meV/Å and the enthalpies converge to better than 1 meV/atom (lower than a chemical accuracy of 1 kcal/mol, ie., 40 meV/atom). The k-grid spacing of 0.20 Å$^{-1}$ is employed following the Monkhorst-Pack scheme. The semi-empirical van der Waals (vdW) interactions are also taken into account by using DFT-D2 functional [22].

To explore the possible polymerization of $N_2F_2$ in molecular crystals upon compression, AIMD simulations based on PBE-D2 calculations at constant pressure-temperature conditions (NpT ensemble) were carried out using the Langevin thermostat [23,24]. The Brillouin zone integration is carried out utilizing the Gamma scheme. The structural transformation (polymerization process) of *cis*-$N_2F_2$ molecular crystal was investigated using 3×3×3 supercells containing 27 molecular $N_2F_2$ units

(108 atoms). Pre-simulation processing involved subjecting these supercells to AIMD simulations (500–2500 K) at each pressure (0–200 GPa) with distinct steps to disrupt symmetry and accelerate equilibration. These processed structures served as initial configurations for subsequent simulations, with the 0 GPa configuraton illustrated in Supporting Information Fig. S1(c). NpT AIMD simulation run for 10 ps with a timestep of 1 fs.

The calculations of the Mean Squared Displacement (MSD) and the diffusion coefficient based on MSD are both derived from the results of AIMD simulations [25]. The gravimetric energy density, detonation velocity, and detonation pressure were calculated using the EXPLO5 software [26]. VASPKIT [27] is also adopted in our work.

The decomposition enthalpy $\Delta H$ is defined as:

$$\Delta H(NF + Al) = \frac{6H(NF) + 2H(Al) - 2H(AlF_3) - 3H(N_2)}{14} \quad (1)$$

Here, $H$(NF), $H$(Al), $H$(AlF$_3$), and $H$(N$_2$) are the enthalpy per formula unit of solid *Cmca* polymerized NF compound, *Fm-3m* Al, *R-3c* AlF$_3$, and the *α*-N$_2$ phase, respectively.

**RESULTS AND DISCUSSION**

It has been predicted that the ground-state NF phase crystallizes in the *Cmca* space group (space group SG=64, Z=8) [16]. Its crystal structure contains infinite covalent chains -(NF)$_n$-. As shown in Fig. 1(a), a zigzag polynitrogen chain is formed by N(sp)$^3$ atoms, with each N atom bearing a terminal fluorine group. This phase is predicted to serve as a HEDM, exhibiting thermodynamically and dynamically stable characteristics within a pressure range of 120 to 200 GPa at 0 K. Furthermore, there are no imaginary frequencies in its phonon spectrum at 0 GPa and 0 K, indicating that this metastable phase constitutes a local minimum on its potential energy surface (PES). The potential for recovering this predicted high-pressure material at atmospheric pressure is especially important. However, the identification of suitable precursor(s) and the establishment of the formation mechanism for the formation of the NF-*Cmca* phase remain challenging.

As shown in Fig. 1(a), the building block of the one-dimensional (1D) poly-NF chain is the difluoride dinitrogen ($N_2F_2$) unit. It can be hypothesised that the precursor of poly-NF chain may be the experimental $N_2F_2$ molecular compound [28,29], which is an unsaturated species with the potential for polymerisation upon compression. The $N_2F_2$ molecule exists in two structural forms: *cis* and *trans* configurations. However, when compared to its *trans* form, the *cis*-isomer evinces greater structural similarity to the NF-*Cmca* phase. Moreover, *cis*-$N_2F_2$ is more stable than *trans*-$N_2F_2$, i.e. 50 meV/$N_2F_2$, i.e. 5 kJ/mol at the PBE level of theory, in perfect agreement with previous ab initio studies [29]. We have therefore chosen a crystalline model containing molecular unsaturated $N_2F_2$ units in the *cis* configuration, as shown in Fig. 1(b). The details are provided in the Supporting Information.

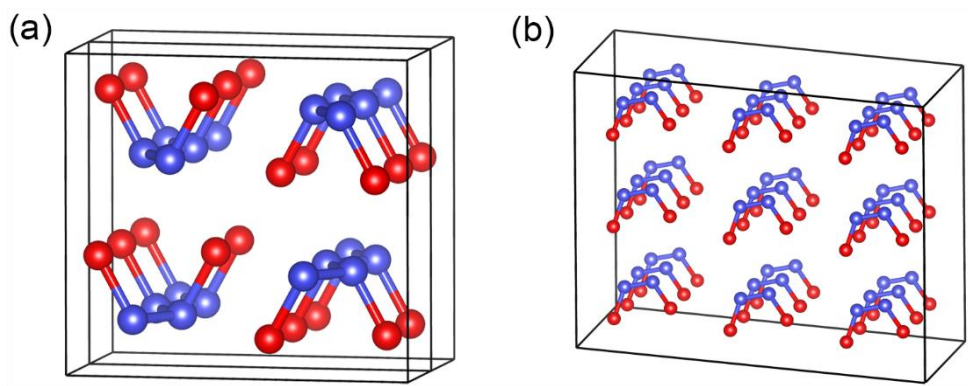

Fig. 1. (a) The 2x1x1 supercell of NF-*Cmca* phase at 120 GPa and 0 K. (b) The 3x3x3 supercell of selected cis-$N_2F_2$ molecular crystal at 0 GPa and 0 K. Blue and red balls represent N and F atoms, respectivelty.

NpT AIMD simulations are performed at pressures ranging from 0 to 200 GPa and temperatures from 0 to 3000 K. The Fig. 2(a) illustrates the degree of polymerization of *cis*-$N_2F_2$ at the end of each simulation.

Below 60 GPa, $N_2F_2$ maintains its molecular form. At 90 GPa and 1900 K, significant polymerization behavior is observed, i.e. finite $(NF)_n$ chains are observed. In the range of 140-200 GPa and around 1900 K, almost complete poly-NF chain is observed. For instance, at 140 GPa, 2000 K, as well as at 160 GPa, 1800 K, and 1900 K, the degree of polymerization reaches 100%. Previous results have suggested that the poly-NF chain is energetically favourable in the range of 120-200 GPa, but at 0 K [16]. Our NpT AIMD results are very consistent with this, showing that the temperature must be lower than 2000 K. Figs. 2(b)-(c) illustrate the results of the MSD at 30 and 120 GPa, showing non-polymerization at 30 GPa and polymerization at 120 GPa. It can be observed that at 30 GPa, as the temperature rises, the MSD gradually increases. On the other hand, at 120 GPa, the MSD shows a pattern that initially increases, then decreases, and increases again with temperature, providing further evidence that polymerisation is taking place. The MSD results at other pressures, shown in Supporting Information Fig. S2, show similar trends.

Previous studies on oxygen-based systems established a diffusion coefficient threshold of $1.0 \times 10^{-5}$ cm²/s as a reliable melting criterion [25]. We extend this standard to NF compounds, given the periodic similarities between N and O atoms. These elements exhibit comparable atomic radii and molar masses, which facilitate similar lattice dynamics. To verify the feasibility, we calculated the diffusion coefficients of configurations at the onset of exhibiting disordered lattice under various pressure conditions. At 0 GPa, as the temperature increases, the configuration begins to exhibit disorder at 600 K, with a diffusion coefficient of $1.42 \times 10^{-5}$ cm²/s. Similarly, at 130 GPa and 2100 K, the configuration shows a diffusion coefficient of $9.50 \times 10^{-6}$ cm²/s. Hence, the diffusion coefficient of $1.0 \times 10^{-5}$ cm²/s can also be considered an appropriate criterion for determining the melting point of NF compound. The corresponding melting curve is depicted in the intermediate region between the pink and brown areas.

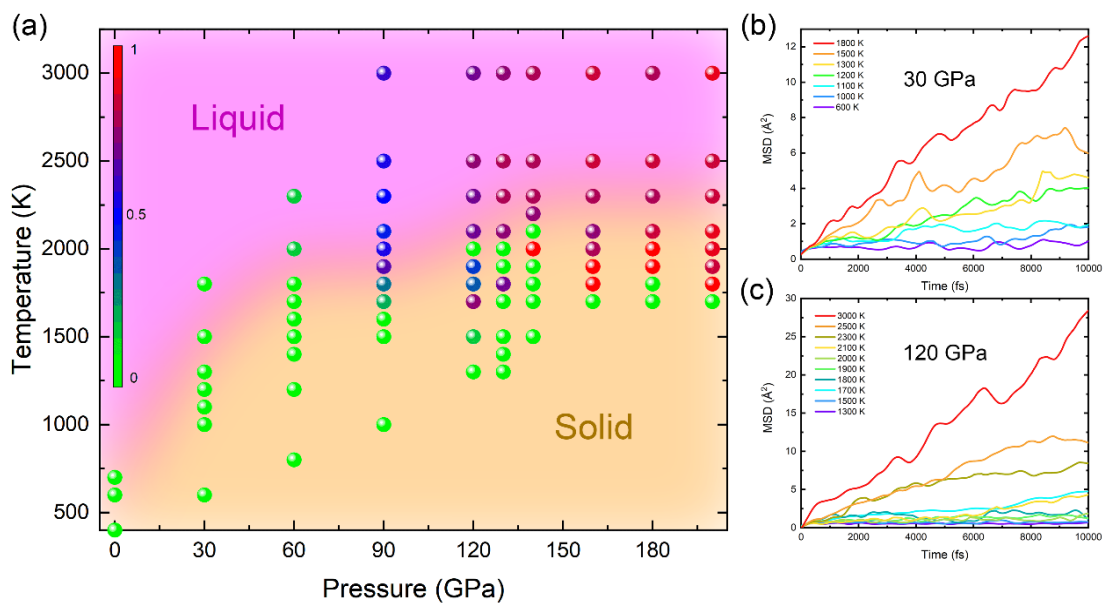

Fig. 2. Dynamical behaviors of $N_2F_2$ molecular at different temperatures and pressures. (a) Phase diagram of NF compound. The pink area represents the melting region, while the orange area represents the solid region. The depth of ball color indicates the degree of polymerization, increasing from green to blue and then to red, with 100% being the highest. (b) MSD at 30 GPa. (c) MSD at 120 GPa. The degree of polymerization was calculated following: Each nitrogen atom bonded to two neighboring nitrogen atoms as a value of 1. The degree of polymerization is the ratio of these counted nitrogen atoms to the total nitrogen atoms in the structure [30].

The structural evolution of $N_2F_2$ units during AIMD simulations at 160 GPa and different temperatures (1600, 1700, and 1800 K) is depicted in Fig. 3(a). For T ≤ 1700 K, the pair correlation function $g(r)$ of N atoms exhibits only one nearest neighbor peak, indicating that $N_2F_2$ remains in a molecular form. At 1800 K, two distinct peaks are observed, demonstrating the polymerization of $N_2F_2$ molecules. Furthermore, the position of the nearest neighbor peak shifts from 1.18 Å to 1.26 Å, suggesting that the bond length is elongated due to an increase in the coordination number of N atoms (double to single bond character, at 160 GPa). Fig. 3(b) shows the dynamic behavior at 1800 K, with a sharp decline in $N_2F_2$ molecules (3500–5000 fs) alongside a rapid increase in poly-NF chain fraction. When the temperature is increased to 3000 K, as depicted in Fig. 3(a), the positions of the two peaks remain unchanged. However, they show a significant broadening, indicating the melting of the poly-NF chains.

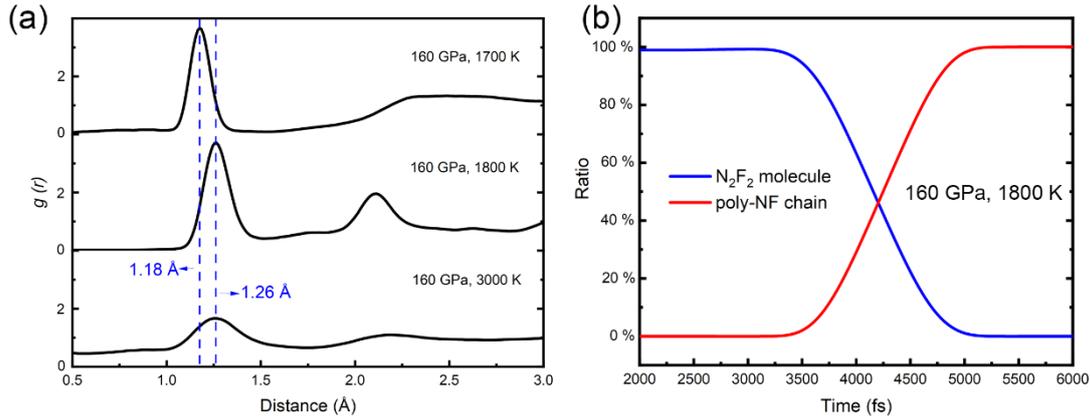

Fig. 3. Dynamical behaviors of $N_2F_2$ molecular at 160 GPa. (a) N-N pair correlation function $g(r)$ at 160 GPa and different temperatures (1700, 1800, and 3000 K). (b) Evolution of the ratio of $N_2F_2$ molecules and polymerized products with time in AIMD simulation at 160 GPa and 1800 K.

The final configurations, obtained from the AIMD simulations at 160 GPa and various temperatures, are relaxed at 160 GPa and 0 GPa, and their enthalpies are compared, as shown in Fig. 4(a). The configuration exists in a molecular form at 1700 K, adopting a polymerized form under other temperatures that melts at T ≥ 2000 K. The configurations of 1700-1900 K are illustrated in Figs. 4(b)-(d). It is observed that the polymerized forms have lower energy compared to the molecular forms under the relaxation of 160 GPa, especially for configurations obtained at 1800 K and 1900 K, where they exhibit fully polymerization and significantly lower energy than others. Therefore, these two configurations are energetically favorable under high pressure conditions. Furthermore, for configuration at 1900 K, due to the higher temperature surpassing the energy barrier, some chains undergo flipping. This flipped configuration has lower energy and higher density compared to the non-flipped one. When they are quenched to 0 GPa for further relaxation, they become to be metastable. For instance, the energies of configurations at 1800 K and 1900 K are -516.24 and -516.28 eV, significantly higher than the energy of the $N_2F_2$ molecular forms (-552.5787 eV). Therefore, the NF chains can easily be polymerized under extreme conditions, and when the pressure is reduced to 0 GPa, they are suitable to be used as an HEDM.

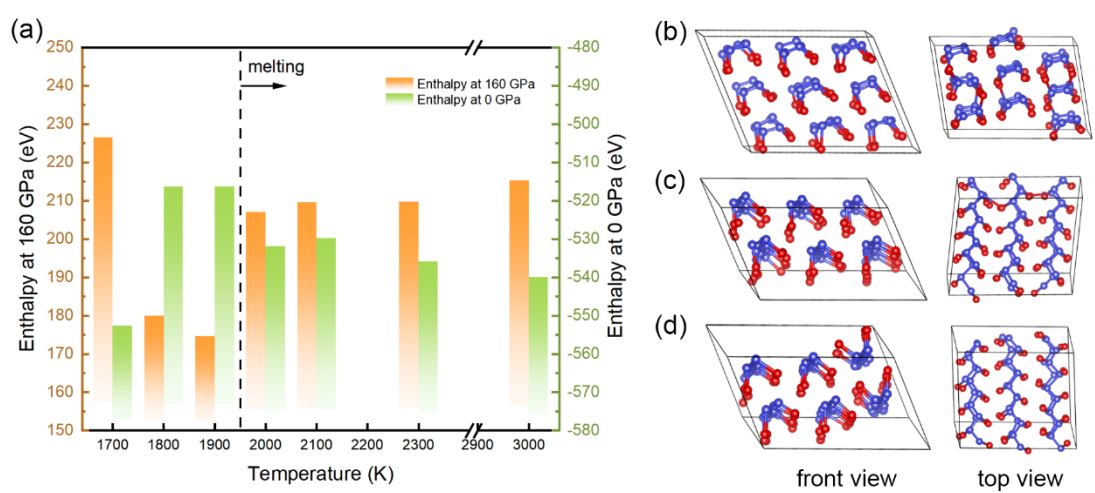

Fig. 4. (a) Enthalpies of relaxed configurations at 160 GPa and 0 GPa were calculated based on the final structures obtained from AIMD simulations conducted at 160 GPa across various temperatures. (b)-(d) Final configurations after the 10 ps AIMD simulations at 160 GPa and (b) 1700 K, (c) 1800 K, and (d) 1900 K. For all structures, blue balls represent N atoms, while red spheres represent F atoms.

As shown in Supporting Information Fig. S3, the average N-N bond length of *cis*-$N_2F_2$ at 0 GPa and 1700 K is 1.27 Å, showing a distinct double N=N bond character [31]. In contrast, the averaged NN interatomic distance of the polymerized chains at 1800 K and 0 GPa is 1.40 Å, a value consistent with the expected single bond character along the poly-NF chains.

Additionally, when the T > 1900 K, the polymerized forms begin to melt. These configurations, under the studied conditions, have similar degree of polymerization, energies, and average N-N bond lengths. However, As a HEDM, it is better to avoid melting during actual polymerization processes because their energies at 0 GPa are lower than the un-melted polymerized forms.

The stability of the poly-NF chains at ambient pressure was investigated. A 10 ps AIMD simulation was conducted to directly quench the configuration from 160 GPa, 1800 K to 160 GPa, 300 K, followed by a 3 ps AIMD simulation at each pressure point during the decompression process, with decompression intervals of 20 GPa. Then, 10 ps AIMD simulations were performed on the quenched configurations at 0 GPa, 300 K and corresponding MSD and average bond lengths variation are shown in Fig. 5. The MSD values fluctuate within a range not exceeding 1.8 Å$^2$, and the variations in the N-N and N-F bond lengths are also very small. For instance, the maximum variation in the N-N bond length is only 5.57% at 300 K. Importantly, the structure is maintained after the AIMD simulations, suggesting that the poly-NF chains are stable at ambient conditions.

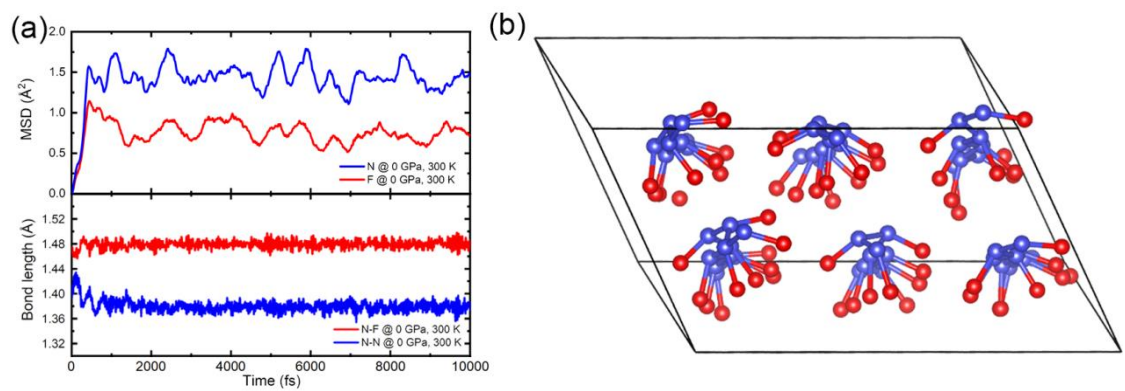

Fig. 5. Dynamics behaviors of poly-NF chains at 0 GPa, 300 K. (a) MSD, average N-F bond length, and average N-N bond length various at time. (b) Final configuration after 10 ps AIMD simulations of 0 GPa, 300 K. Blue spheres represent N atoms, while red spheres represent F atoms.

The detonation performance of the poly-NF chains as a standalone reactant that produces $NF_3$ and $N_2$ has been studied [16]. In this work, the decomposition enthalpy of the *Cmca* NF compound in the presence of solid Al as a reducing agent, relative to $AlF_3$ and $N_2$ at ambient pressure and 0 K is calculated. The value of 2.44 eV/atom indicates that the NF-*Cmca* phase can release a significant amount of energy when reacting with Al.

For the calculation of gravimetric energy density ($E_d$), to account for the transition from solid to gas, we added a correction based on the chemical potential. The energy reduction from solid $N_2$ to gas is estimated by adding the calculated chemical potential at 298 K to the 0 K enthalpy, which is 0.50 eV/$N_2$, obtained from the thermochemical tables [32]. The calculated $E_d$ and explosive performance of NF+Al is shown in Fig. 6, where previous reported experimental values of $\varepsilon$-CL-20, TNT, and HMX are included for comparation [33–35]. Calculations revealed that $E_d$ of NF+Al is estimated to be approximately 13.55 kJ/g, which is more than 3 times as high as that of TNT (4.30 kJ/g) and 2-3 times as high as that of HMX (5.70 kJ/g). The volumetric energy density ($E_v$) calculations further highlight its superior performance, with a value of 35.90 kJ/cm$^3$. This value is over 5 times greater than that of TNT (7.05 kJ/cm$^3$) and over 3 times greater than HMX (10.77 kJ/cm$^3$). Furthermore, the detonation velocity ($V_d$) and detonation pressure ($P_d$) of NF+Al are 11.85 km/s and 750.42 kbar, respectively. As a result, NF+Al is a HEDM with high performance.

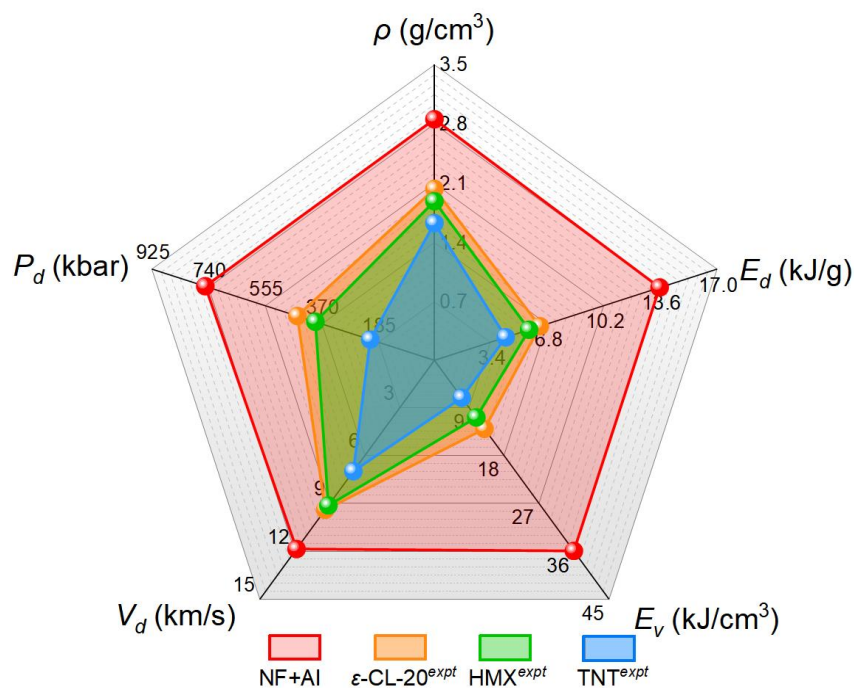

Fig. 6. The calculated density $\rho$, gravimetric energy density $E_d$, volumetric energy density $E_v$, detonation velocity $V_d$, and detonation pressure $P_d$ of NF+Al. For comparison, experimental results for $\varepsilon$-CL-20, TNT, and HMX explosives are also listed. Herein, superscript expt represents the experimental data.

**CONCLUSION**

Based on first-principles calculations and AIMD simulations, we have studied the dynamic behaviour of $cis$-$N_2F_2$ molecules under different temperature (0-3000 K) and pressure (0-200 GPa) conditions, characterised the temperature-pressure phase diagram of the NF compound and determined its melting temperature.

Our NpT AIMD modelling shows that polymerisation of $cis$-$N_2F_2$ molecules is possible at pressures above 90 GPa and temperature around 2000 K. This poly-NF chains can be quenched to ambient conditions and has a temperature stability of 300 K.

When Al is used as a reducing agent, the NF+Al composite formed by $Cmca$-type NF and Al has an impressive gravimetric energy density of up to 13.55 kJ/g. This value is 3-4 times higher than that of TNT and significantly higher than the 9.7 kJ/g of cg-N.

**ACKNOWLEDGEMENTS**

This work is supported by the National Natural Science Foundation of China (NSFC) under Grant of U2030114, and CASHIPS Director's Fund (Grant No. YZJJ202207-CX). The calculations were partly performed in Center for Computational Science of CASHIPS, the ScGrid of Supercomputing Center and Computer Network Information Center of Chinese Academy of Sciences, and the Hefei Advanced Computing Center. GF thanks the Region Nouvelle Aquitaine (France) and the European Union (ERDF).

# Unsaturated Dinitrogen Difluoride under Pressure: toward high-Energy Density Polymerized NF Chains

Guo Chen[1,2], Ling Lin[3], Chengfeng Zhang[1,2], Jie Zhang[1,*], Gilles Frapper[4] and Xianlong Wang[1,2*]

[1]*Key Laboratory of Materials Physics, Institute of Solid State Physics, HFIPS, Chinese Academy of Sciences, Hefei 230031, China*

[2]*University of Science and Technology of China, Hefei 230026, China*

[3]*State Key Laboratory of Fluorine and Nitrogen Chemistry and Advanced Materials, Shanghai Institute of Organic Chemistry, Chinese Academy of Sciences, Shanghai 200032, China*

[4]*Applied Quantum Chemistry group, E4, IC2MP, UMR 7285 Poitiers University, CNRS, 4 rue Michel Brunet TSA 51106, 86073 Poitiers Cedex 9, France.*

―――――

[*]Author to whom all correspondence should be addressed: xlwang@theory.issp.ac.cn,

jzhang@theory.issp.ac.cn

# CONTENT

## S1 Introducing of *cis*-N$_2$F$_2$ Molecular Crystal

The *cis*-N$_2$F$_2$ molecular crystal structure (OQMD ID: 1947574) with space group *P1*, containing one molecular unit per unit cell, can be retrieved from the Open Quantum Materials Database. After relaxation to 0 GPa, its enthalpy value was determined to be -5.0648 eV/atom. In contrast, a self-designed isostructural crystal with space group *P1* shows superior energy characteristics (-5.0750 eV/atom) under the same pressure conditions. Structural differences between the two configurations are illustrated in Supporting Information Fig. S1.

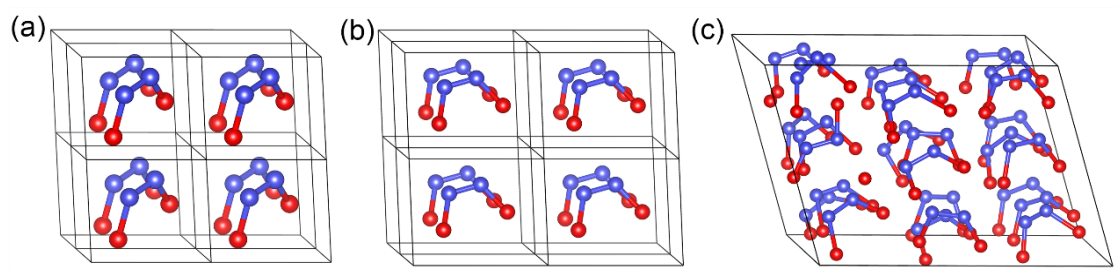

Fig. S1. Structural configurations of *cis*-N$_2$F$_2$ molecular crystals. (a) Structure from OQMD (ID: 1947574). (b) Self-designed isostructural variant. (c) Initial configuration for AIMD simulations at 0 GPa.

Table S1 Calculated energies (eV/atom) at 0 GPa and crystallographic parameters (Å, °) of *cis*-N$_2$F$_2$ molecular crystal from OQMD and self-designed.

|  | Space group | Energy (eV/atom) | Crystallographic parameters (Å, °) |
| --- | --- | --- | --- |
| *cis*-N$_2$F$_2$ from OQMD | *P1* | -5.0648 | a=3.76, b=3.98, c=3.95; N-N=1.2136, N-F=1.4221 |
| *cis*-N$_2$F$_2$ from self-designed | *P1* | -5.0750 | a=4.98, b=3.99, c=3.08; N-N=1.2163, N-F=1.4224 |

S3

## S2 MSD of NpT Simulations

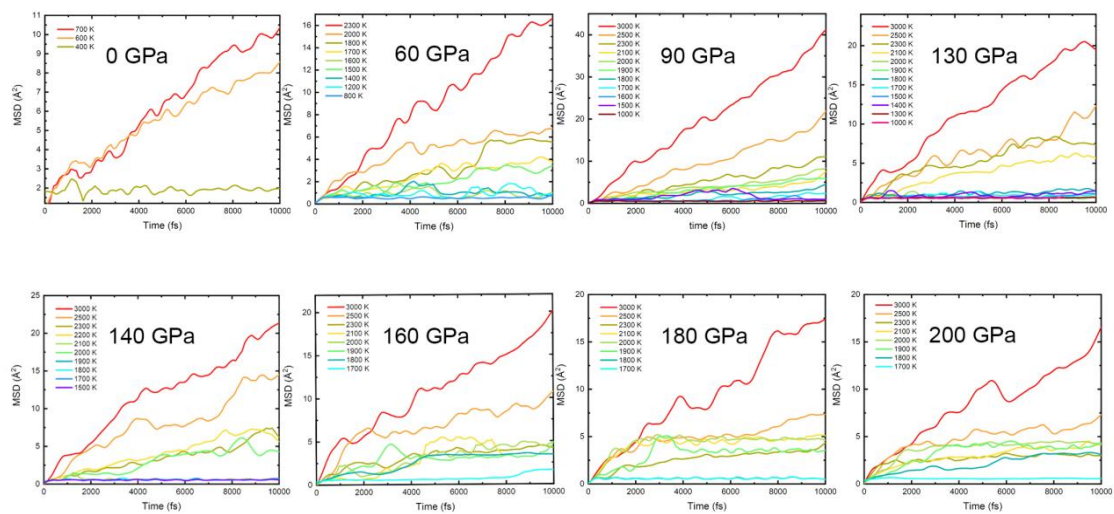

Fig. S2. MSD of $N_2F_2$ molecular crystal at different at different temperatures and pressures.

## S3 N-N Bond Lengths and Degree of Polymerization in the Final Configurations Obtained from AIMD Simulations at 160 GPa

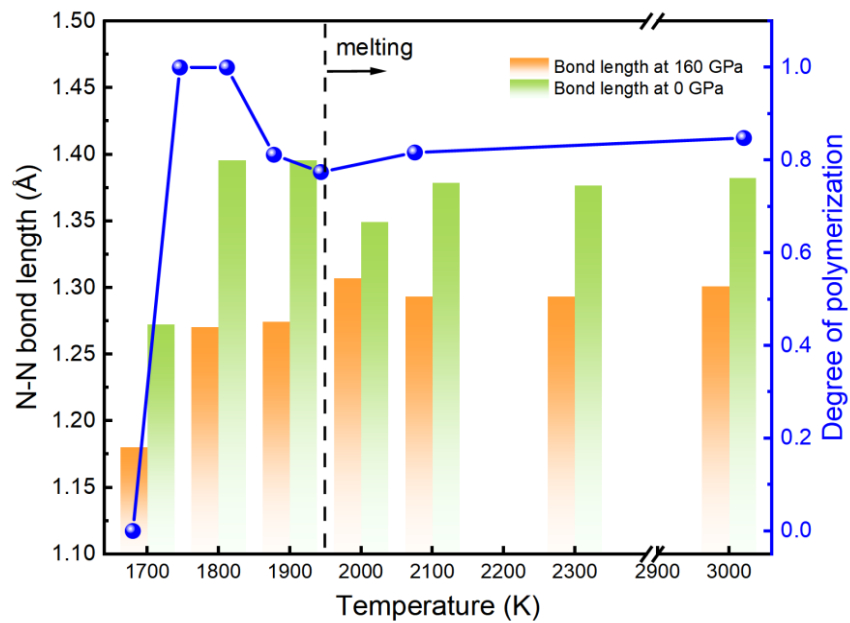

Fig. S3. N-N Bond lengths and degree of polymerization for relaxed configurations at 160 GPa and 0 GPa were calculated based on the final structures obtained from AIMD simulations conducted at 160 GPa across various temperatures.